\def\pp{\noindent\parshape 2 0truecm 8.5 truecm 0.8truecm 7.7truecm}
\def\rk#1;#2;#3;#4;#5 {\pp\addtocounter{enumi}{1}{\bf (\arabic{enumi})} ``#1", #2, {\it #3}, {\bf #4}, #5}
\def\ro{\pp\addtocounter{enumi}{1}{\bf (\arabic{enumi})}~}

\def\beq#1{\begin{equation}\label{#1}}
\def\eeq{\end{equation}}
\def\beqa#1{\begin{eqnarray}\label{#1}}
\def\eeqa{\end{eqnarray}}
\def\eq#1{equation~(\ref{#1})}

\def\fig#1{Figure~\ref{#1}}
\def\Fig#1{Figure~\ref{#1}}
\def\tab#1{Table~\ref{#1}}
\def\Tab#1{Table~\ref{#1}}

\def\Sec#1{Section~\ref{#1}}
\def\Sec#1{Section~\ref{#1}}

\def\ie{{\frenchspacing\it i.e.}}
\def\eg{{\frenchspacing\it e.g.}}

\def\rms{{\frenchspacing r.m.s.}}

\def\spose#1{\hbox to 0pt{#1\hss}}
\def\simlt{\mathrel{\spose{\lower 3pt\hbox{$\mathchar"218$}}
     \raise 2.0pt\hbox{$\mathchar"13C$}}}
\def\simgt{\mathrel{\spose{\lower 3pt\hbox{$\mathchar"218$}}
     \raise 2.0pt\hbox{$\mathchar"13E$}}}
\def\simpropto{\mathrel{\spose{\lower 3pt\hbox{$\mathchar"218$}}
     \raise 2.0pt\hbox{$\propto$}}}

\def\rms    {{\it rms~}}
\def\ra{{\rm ra}}
\def\dec{{\rm dec}}

\def\rhat{\widehat{\bf r}}
\def\xhat{\widehat{\bf x}}
\def\nbf{\bf n}
\def\xbf{\bf x}
\def\ybf{\bf y}
\def\Abf{\bf A}
\def\Nbf{\bf N}
\def\Atbf{{\bf A}^{\rm T}}
  
\def\expec#1{\langle#1\rangle}

\def\A{{\bf A}}
\def\C{{\bf C}}
\def\I{{\bf I}}
\def\LL{{\bf\Lambda}}
\def\N{{\bf N}}
\def\NN{{\bf\Sigma}}
\def\P{{\bf P}}
\def\Pt{\tilde{\bf P}}
\def\R{{\bf R}}
\def\a{{\bf a}}
\def\at{\tilde{\bf a}}
\def\n{{\bf n}}

\def\xhat{\hat{\bf x}}
\def\y{{\bf y}}
\def\z{{\bf z}}
\def\zero{{\bf 0}}

\def\jth{j^{\rm th}}

\def\mstar{m_*}
\def\npix{n_{\rm pix}}
\def\tr{{\rm tr}\>}
\def\mK{{\rm mK}}
\def\uK{\mu{\rm K}}

\documentclass[useAMS,usenatbib,usegraphicx,numcite]{mn2e}
\usepackage{times}

\title[A Model of Diffuse Galactic Radio Emission]
      {A Model of Diffuse Galactic Radio Emission\\from 10 MHz to 100 GHz}

\author[de Oliveira-Costa et al.]{Ang{\'e}lica de Oliveira-Costa$^1$, Max Tegmark$^1$, B.M. Gaensler$^2$, 
\newauthor Justin Jonas$^3$, T.L. Landecker$^4$, Patricia Reich$^5$  \\
$^1$MIT Kavli Institute \& Dept.~of Physics, Massachusetts Institute of Technology, Cambridge, MA 02139, USA \\
$^2$School of Physics A29, The University of Sydney, NSW 2006, Australia \\
$^3$Department of Physics \& Electronics, Rhodes University, Grahamstown 6140, South Africa \\
$^4$Dominion Radio Astrophysical Observatory, National Research Council, P.O. Box 248, Penticton, B.C., Canada \\
$^5$Max-Planck-Institut f\"ur Radioastronomie, Auf dem H\"ugel 69, D-53121 Bonn, Germany \\
}

\begin{document}

\date{Accepted ---; Received ---; in original form ---.}

\pagerange{\pageref{firstpage}--\pageref{lastpage}}

\pubyear{2008}

\maketitle

\label{firstpage}

\begin{abstract}
Understanding diffuse Galactic radio emission is interesting both in its own right and for 
minimizing foreground contamination of cosmological measurements. 
Cosmic Microwave Background experiments have focused on frequencies $\simgt 10\>$GHz,
whereas 21 cm tomography of the high redshift universe will mainly focus on 
$\simlt 0.2\>$GHz, for which less is currently known about Galactic emission.
Motivated by this, we present a global sky model derived from all publicly available 
total power large-area radio surveys, digitized with optical character 
recognition when necessary and compiled into a uniform format, as well as the new Villa Elisa data 
extending the 1.42 GHz map to the entire sky.
We quantify statistical and systematic uncertainties in these surveys by comparing them
with various global multi-frequency model fits. We find that a principal component based model
with only three components can fit the 11 most accurate data sets 
(at 10, 22, 45 \& 408 MHz and 1.42, 2.326, 23, 33, 41, 61, 94 GHz)
to an accuracy around 1\%-10\% depending on frequency and sky region.
Both our data compilation and our software returning a predicted all-sky map at any frequency 
from 10 MHz to 100 GHz are publicly available at 
{\bf http://space.mit.edu/home/angelica/gsm}. 
\end{abstract}



\setcounter{footnote}{0}

\section{Introduction}

\begin{figure*} 
\includegraphics[width=18.3cm]{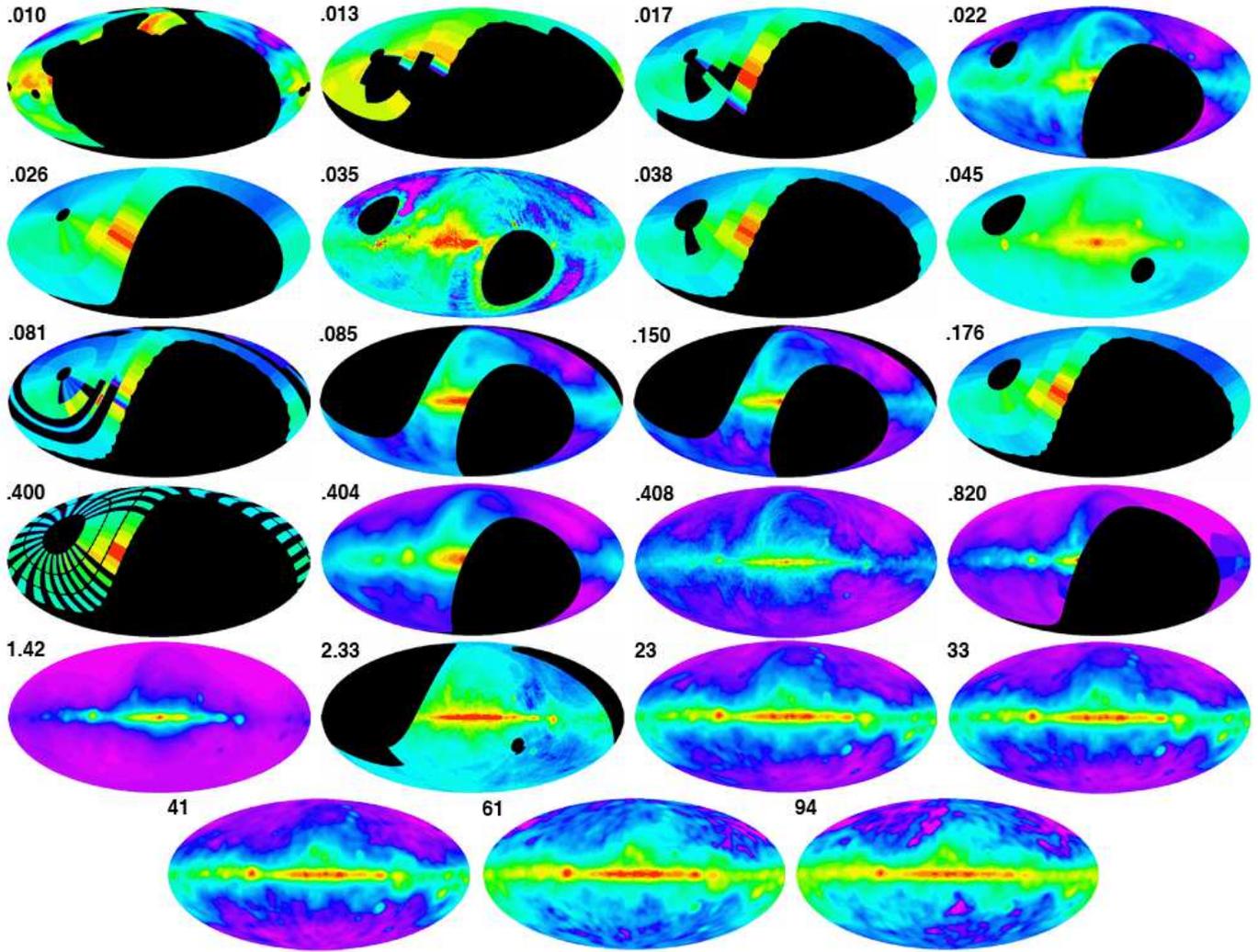}
\caption{
The maps show (from left to right, top to bottom) the  
 0.010   GHz \protect\cite{caswell}, 
 0.0135  GHz \protect\cite{bridle}, 
 0.0175  GHz \protect\cite{bridle}, 
 0.022   GHz \protect\cite{roger}, 
 0.026   GHz \protect\cite{turtle}, 
 0.0345  GHz \protect\cite{shankar}, 
 0.038   GHz \protect\cite{turtle}, 
 0.045   GHz \protect\cite{alvarez},
 0.0815  GHz \protect\cite{bridle},
 0.085   GHz \protect\cite{land85}, 
 0.150   GHz \protect\cite{land85},  
 0.176   GHz \protect\cite{turtle},  
 0.400   GHz \protect\cite{turtle}, 
 0.404   GHz \protect\cite{pauliny},  
 0.408   GHz \protect\cite{haslam2},  
 0.820   GHz \protect\cite{berk},  
 1.42    GHz \protect\cite{reichreich,reichreich2,reich01} and
 2.326   GHz \protect\cite{jonas} surveys, and 
the CMB-free WMAP foreground maps at 23, 33, 41, 61 and 94 GHz 
\protect\cite{mapforegs,multipoles,mapforegs2,thiswork}. 
}
\label{AllMapsFig}
\end{figure*}

There has been a great deal of interest in mapping diffuse Galactic radio emission, both to better 
understand our Galaxy and to clean out foreground contamination from Cosmic Microwave Background 
(CMB) maps. For reviews of such issues, see, \eg, 
from  \cite{wiener} to \cite{saha},  
and references therein.
Because much of this work was both motivated by the CMB and based on maps from CMB experiments, 
it has focused primarily on frequencies above 10 GHz. Now that Neutral Hydrogen Tomography (NHT) 
is emerging as a promising cosmological tool where we can map the high-redshift universe 
three-dimensionally via the redshifted 21 cm line
-- see, \eg, from  \cite{rennan04a} to \cite{asantha04} -- 
it is timely to extend these efforts down to lower frequencies. The redshift range $7\simgt z\simgt 50$ where 
NHT may be feasible corresponds to the frequency range $30-180$ MHz. 
Indeed, accurate foreground modeling is arguably even more important for NHT than for CMB: whereas 
unpolarized CMB fluctuations dominate over foregrounds for the most favorable frequencies and sky 
directions, and the situation for polarized CMB fluctuations is only 1-2 orders of magnitude worse, 
the cosmological neutral hydrogen signal is perhaps four orders of magnitudes smaller than the 
relevant foregrounds -- see, \eg, from  \cite{miguel03} to \cite{FurlanettoReview}.

In due time, NHT experiments such as LOFAR \cite{lofar03,lofar03b,zaroubi04}, MWA \cite{miguel03,bowman05} 
and 21CMA \cite{peterson05} will produce accurate low-frequency maps of Galactic emission much like CMB 
experiments have above 10 GHz. Even now, however, it is important to have a model for how this emission 
varies across the sky and across the NHT-relevant frequency band, because this helps optimize experimental 
design, scan strategy and data analysis pipelines to maximize the scientific return on its investment.
Even as new maps are made of small patches of sky, it is valuable to have a pre-existing 
global sky model to be able to quantify and mitigate contamination from distant sidelobes.

\begin{figure} 
\includegraphics[width=9cm]{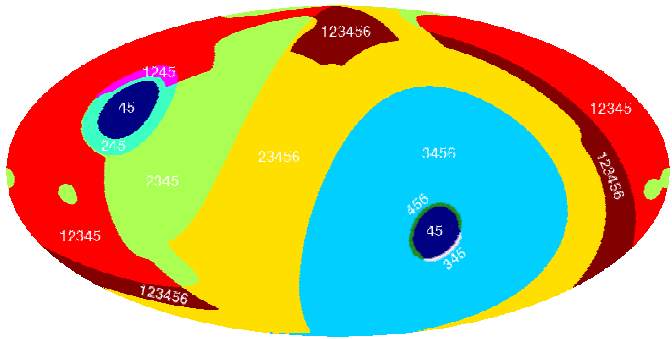}
\caption{Sky coverage/overlap, in Galactic coordinates, of the six key maps below the WMAP frequencies:
         the numbers from $1$ to $6$ correspond to 10, 22, 45, 408, 1420 and 2326 MHz, respectively.}
\label{SurveysSkyCovarageFig}
\end{figure}

As described in \Sec{DataSec}, the radio astronomy community has produced a large number of sky surveys 
over the years, many of which are directly relevant to the NHT frequency range. However, most of them have 
never been used by cosmology community, because of various challenges:
many are not publicly available in electronic form, and they are also hard to combine because they differ 
in sky coverage, angular resolution, pixelization and quality. 
Instead, a common approach among cosmologists has been to simply extrapolate the 408 MHz Haslam map 
\cite{haslam2} to lower frequencies, thus ignoring any spectral variation across the sky. As we will see, 
significantly better accuracy can be attained by modeling that includes additional data sets.
The goal of the present paper is to collect, standardize and model this large body of radio data to make 
it more useful to the cosmology community.

The rest of this paper is organized as follows. In \Sec{DataSec}, we describe how we compile all publicly 
available total power large-area radio surveys of which we are aware, digitizing them with optical character 
recognition when necessary, and converting them into a uniform format. In \Sec{MethodSec}, we compare 
different methods for constructing a global sky model from this data that covers the entire sky and the 
entire frequency range. In \Sec{ResultsSec}, we present the results of our modeling, quantify the accuracy 
of our best model, and briefly comment on implications for the physics underlying this emission.


\def\h{{$^{\rm h}$}}
\def\d{{$^{\circ}$}}
\def\ll{{$<$}}
\def\g{{$>$}}
\def\ra{{$\alpha$}}
\def\dec{{$\delta$}}
\def\app{{$\approx$}}
\def\elle{{$\ell$}}
\def\bb{{$b$}}
\def\modbb{{$|b|$}}
\begin{table*}\label{mytable1}
\begin{center}
\caption{Available total power radio surveys.}
\bigskip
{\footnotesize
\begin{tabular}{l|c|c|cc|c|r}
\hline
\hline
\multicolumn{1}{ l|}{Ref}          &
\multicolumn{1}{|c|}{$\nu$}        &
\multicolumn{1}{|c|}{FWHM}         &
\multicolumn{2}{|c|}{Region}       &
\multicolumn{1}{|c|}{Observatory}  &
\multicolumn{1}{|r}{Status}        \\
&[GHz] &[\d]  &RA &DEC & &              \\
\hline
\hline  
 \protect\cite{alex}       &0.00393  &60      & 00\h \ll \ra   \ll  24\h    &$-$60\d \ll \dec \ll$+$60\d                &RAE-1		     &D\\
 \protect\cite{alex}       &0.00655  &60      & 00\h \ll \ra   \ll  24\h    &$-$60\d \ll \dec \ll$+$60\d                &RAE-1		     &D\\ 
 \protect\cite{caswell}    &0.010    &2.6x1.9 & 00\h \ll \ra   \ll  16\h    &$-$06\d \ll \dec \ll$+$74\d	        &DRAO, CAN	     &A\\ 
 \protect\cite{cane2}      &0.010    &5       & 00\h \ll \ra   \ll  24\h    &$-$65\d \ll \dec \ll$+$90\d	        &DRAO(CAN),AUS       &D\\ 
 \protect\cite{hamilton}   &0.0102   &4x5     & 00\h \ll \ra   \ll  24\h    &$-$65\d \ll \dec \ll$-$02\d	        &AUS		     &D\\
 \protect\cite{bridle}     &0.0135   &53x12   & 00\h \ll \ra   \ll  24\h    &(16,35,52,69)\d		                &ENG		     &C\\ 
 \protect\cite{ellis}      &0.0165   &1.5     & 00\h \ll \ra   \ll  24\h    &$-$90\d \ll \dec \ll~~  00\d	        &Tasmania, AUS       &D\\ 
 \protect\cite{bridle}     &0.0175   &12x17   & 00\h \ll \ra   \ll  24\h    &(16,35,52,69)\d		                &ENG		     &C\\ 
 \protect\cite{roger}      &0.022    &1.1x1.7 & 00\h \ll \ra   \ll  24\h    &$-$28\d \ll \dec \ll $+$80\d	        &DRAO, CAN	     &A\\ 
 \protect\cite{turtle}     &0.026    &15x44   & 00\h \ll \ra   \ll  24\h    &(20,40,60)\d		                &ENG		     &C\\ 
 \protect\cite{mathewson}  &0.030    &11      & 00\h \ll \ra   \ll  24\h    &$-$90\d \ll \dec \ll~~  00\d	        &Parkes, AUS	     &D\\  
 \protect\cite{cane}       &0.030    &11      & 00\h \ll \ra   \ll  24\h    &$-$90\d \ll \dec \ll $+$90\d	        &		     &D\\  
 \protect\cite{shankar}    &0.0345   &0.4x0.7 & 00\h \ll \ra   \ll  24\h    &$-$30\d \ll \dec \ll $+$60\d	        &GEETEE, IND	     &A\\  
 \protect\cite{milo}       &0.038    &7.5     & 00\h \ll \ra   \ll  24\h    &$-$25\d \ll \dec \ll $+$70\d	        &Jodrell Bank, ENG   &D\\  
 \protect\cite{turtle}     &0.038    &15x44   & 00\h \ll \ra   \ll  24\h    &(20,40,60)\d		                &ENG		     &C\\  
 \protect\cite{alvarez}    &0.045    &4.6x2.4 & 00\h \ll \ra   \ll  24\h    &$-$90\d \ll \dec \ll $+$19\d	        &CHL		     &B\\  
 \protect\cite{maeda}      &0.045    &3.6x3.6 & 00\h \ll \ra   \ll  24\h    &$+$05\d \ll \dec \ll $+$65\d	        &JPN		     &B\\  
 \protect\cite{bridle}     &0.0815   &12x17   & 00\h \ll \ra   \ll  24\h    &(16,25,30,35,40,52,69)\d                   &ENG		     &C\\  
 \protect\cite{land85}     &0.085    &3.8x3.5 & 00\h \ll \ra   \ll  24\h    &$-$25\d \ll \dec \ll $+$25\d	        &Parkes, AUS	     &C,A\\  
 \protect\cite{bolton}     &0.100    &17      & 00\h \ll \ra   \ll  24\h    &$-$90\d \ll \dec \ll $+$30\d               &AUS		     &D  \\
 \protect\cite{land85}     &0.150    &2.2x2.2 & 00\h \ll \ra   \ll  24\h    &$-$25\d \ll \dec \ll $+$25\d               &Parkes, AUS	     &C,A\\  
 \protect\cite{hamilton2}  &0.153    &2.2     & 00\h \ll \ra   \ll  24\h    &$-$90\d \ll \dec \ll $+$05\d	        &Parkes, AUS	     &D  \\  
 \protect\cite{reber44}    &0.160    &8x6     & 16\h \ll \ra   \ll  22\h    &$-$33\d \ll \dec \ll $+$90\d	        &Wheaton, USA	     &D\\  
 \protect\cite{turtle}     &0.176    &15x44   & 00\h \ll \ra   \ll  24\h    &(20,40,60)\d  	                        &ENG		     &C\\  
 \protect\cite{turtle2}    &0.178    &0.2x4.6 & 00\h \ll \ra   \ll  24\h    &$-$05\d \ll \dec \ll $+$90\d               &ENG		     &D\\  
 \protect\cite{allen}      &0.200    &        & 00\h \ll \ra   \ll  24\h    &$-$90\d \ll \dec \ll $+$45\d	        &Commonwealth, AUS   &D\\  
 \protect\cite{drogue}     &0.200    &16.8    & 00\h \ll \ra   \ll  24\h    &$-$20\d \ll \dec \ll $+$90\d	        &Kieler, GER	     &D\\  
 \protect\cite{turtle}     &0.400    &8.5x6.5 & 00\h \ll \ra   \ll  24\h    &(20,40,60)\d  	                        &ENG		     &C\\  
 \protect\cite{pauliny}    &0.404    &7.5     & 00\h \ll \ra   \ll  24\h    &$-$20\d \ll \dec \ll $+$90\d	        &ENG		     &C\\  
 \protect\cite{haslam2}    &0.408    &0.8     & 00\h \ll \ra   \ll  24\h    &$-$90\d \ll \dec \ll $+$90\d	        &GER, AUS, ENG       &A\\  
 \protect\cite{berk}       &0.820    &1.2     & 00\h \ll \ra   \ll  24\h    &$-$07\d \ll \dec \ll $+$85\d	        &Dwingeloo, NLD      &C,A\\
 \protect\cite{reichreich,reichreich2} 
                           &1.42    &0.6     & 00\h \ll \ra   \ll  24\h     &$-$19\d \ll \dec \ll $+$90\d   		&Stockert, GER       &A\\  
 \protect\cite{reich01} 
                           &1.42    &0.6     & 00\h \ll \ra   \ll  24\h     &$-$90\d \ll \dec\ll$-$10\d     	 	&Villa Elisa, ARG    &B\\  
 \protect\cite{tello07}    &2.3     &2.3x1.9 & 00\h \ll \ra   \ll  24\h     &$-$53\d \ll \dec \ll $+$35\d   		&BRA		     &D\\  
 \protect\cite{jonas}      &2.326   &0.3     & 00\h \ll \ra   \ll  24\h     &$-$83\d \ll \dec \ll $+$32\d   		&Hartebeesthoek, ZAF &B\\  
 \protect\cite{cott,steve}                                &19       &3    & 00\h \ll \ra \ll 24\h  &$-$15\d \ll \dec \ll $+$75\d &Ballon, USA         &B\\ 
 \protect\cite{mapforegs,multipoles,mapforegs2,thiswork}  &23       &0.88 & 00\h \ll \ra \ll 24\h  &$-$90\d \ll \dec \ll $+$90\d &WMAP	             &A\\ 
 \protect\cite{bennett}                                   &31       &7    & 00\h \ll \ra \ll 24\h  &$-$90\d \ll \dec \ll $+$90\d &COBE/DMR	     &A\\ 
 \protect\cite{mapforegs,multipoles,mapforegs2,thiswork}  &33       &0.66 & 00\h \ll \ra \ll 24\h  &$-$90\d \ll \dec \ll $+$90\d &WMAP	             &A\\ 
 \protect\cite{mapforegs,multipoles,mapforegs2,thiswork}  &41       &0.51 & 00\h \ll \ra \ll 24\h  &$-$90\d \ll \dec \ll $+$90\d &WMAP	             &A\\ 
 \protect\cite{bennett}                                   &53       &7    & 00\h \ll \ra \ll 24\h  &$-$90\d \ll \dec \ll $+$90\d &COBE/DMR	     &A\\ 
 \protect\cite{mapforegs,multipoles,mapforegs2,thiswork}  &61       &0.35 & 00\h \ll \ra \ll 24\h  &$-$90\d \ll \dec \ll $+$90\d &WMAP	             &A\\ 
 \protect\cite{bennett}                                   &90       &7    & 00\h \ll \ra \ll 24\h  &$-$90\d \ll \dec \ll $+$90\d &COBE/DMR	     &A\\ 
 \protect\cite{mapforegs,multipoles,mapforegs2,thiswork}  &94       &0.22 & 00\h \ll \ra \ll 24\h  &$-$90\d \ll \dec \ll $+$90\d &WMAP	             &A\\ 
\hline								        		 
\hline
\end{tabular}
}
\label{RadioTable}
\end{center}
\noindent{\small \\
		 A = Publicly available in digital form.\\
		 B = Available on request. \\
		 C = Available as printed table (which we OCRed). \\ 
		 D = Not available in any numerical form.
		 }\\
\end{table*}

\section{Data Sets}  
\label{DataSec}

In order to carry out our analysis, we performed a literature search for large-area total power 
sky surveys in the frequency range 1 MHz to 100 GHz. The result of our search is shown in 
\fig{AllMapsFig} and \tab{RadioTable}. Unfortunately, some of the  surveys shown in 
\tab{RadioTable} are not available in any numerical form. A minority of the surveys 
are publicly available in digital form (and/or) available on request, while many of 
the surveys are available only as printed tables which we converted to digital form
using Optical Character Recognition (OCR).
 
Some of the surveys shown in \fig{AllMapsFig} have a very large angular beam (the 0.0135, 
0.0175, 0.026, 0.038, 0.0815, 0.176, 0.400 and the 0.404 GHz maps), others are undersampled 
(the 0.085 and the 0.150 surveys), the 0.820 GHz survey is smoothed to 5$^{\circ}$ in its 
anti-centre area and, finally, one of them, the 0.0345 GHz map, it is missing large-scale 
structures and has severe striping that makes it unsuitable for use in our analysis. 
Therefore, all analysis presented in this paper is performed using the 0.010, 0.022, 0.045, 
0.408, 1.42, 2.326, 23, 33, 41, 61 and the 94 GHz maps -- \fig{SurveysSkyCovarageFig} 
shows the sky coverage of these different surveys and how they overlap. 
They were all transformed to Galactic coordinates, pixelized using the HealPix RING scheme 
\cite{healpix} with resolution $nside$=512 (which corresponds to $12\times 512^2=3,145,728$ 
equal-area pixels across the sky), and had the CMB component of 2.725 K subtracted. 
For the five 5-year WMAP maps \cite{hinshaw2} used in this analysis, we removed their CMB component as 
described in \cite{mapforegs,mapforegs2}\footnote{
	The new foreground-cleaned 5 year WMAP map and the five foreground-only maps are available on the web site
        {\bf http://space.mit.edu/home/angelica/gsm} (bottom of page). 
	} 
and then converted these maps to antenna temperature. Before performing our main analysis, 
we smooth all maps to a common final angular resolution of 5.1$^{\circ}$.
However, as described below, we also use full-resolution maps for some other purposes.


\section{Methods}
\label{MethodSec}

A wide variety of models of Galactic emission have been used in the literature -- see, \eg, 
\cite{wiener,bouchet,bouchet99,foregpars,giardino02} and references therein, and there are 
many additional popular fitting techniques that are purely statistical in nature and do not 
assume anything about the underlying physics. 
To maximize the utility of the available data sets and all the hard work that observers have 
invested into obtaining them, we explored a wide range of modeling methods before selecting one.
Below we first present the criterion we will use for choosing the best modeling technique, then 
explore a range of methods to select the one that is most useful for our goal. The models we 
compare include physics-inspired fitting functions, power laws, polynominals and splines as
well as principal components.

\subsection{Criteria: accuracy and simplicity}
\label{AccuracyCriterionSec}

In this paper, our main criterion for chosing a method is accuracy. In other words, 
we wish to find the method that most accurately predicts the Galactic emission in any arbitrary 
sky direction and at any frequency between 0.010 to 94 GHz, independently of whether it is 
based on physical assumptions or is ``blind'' and purely statistical. 
In practice, we implement this criterion as follows: for each of the 11 frequencies 
where a high-quality sky map is available, we quantify how accurately a method 
can predict this map by using only information from the other 10 maps.

Since the map used as the ``truth'' in each test may itself have noise and systematic errors, 
this procedure can overestimate the true errors. Moreover, our final Global Sky Model (GSM) 
uses all 11 input maps jointly, not merely 10 at a time, and it is therefore more accurate 
than the model used in the test. For both of these reasons, the accuracy numbers we quote 
later on should be interpreted as conservative worst-case bounds on the actual accuracy.

In addition to accuracy, we also desire simplicity. Specifically, it is valuable if the 
modeling method is simple and transparent enough to allow an analytical understanding of 
how the input determines the output, especially if this makes it possible to characterize 
how noise and systematic errors propagate and affect the statistical properties of the 
resulting model.

\begin{figure} 
\includegraphics[width=9cm]{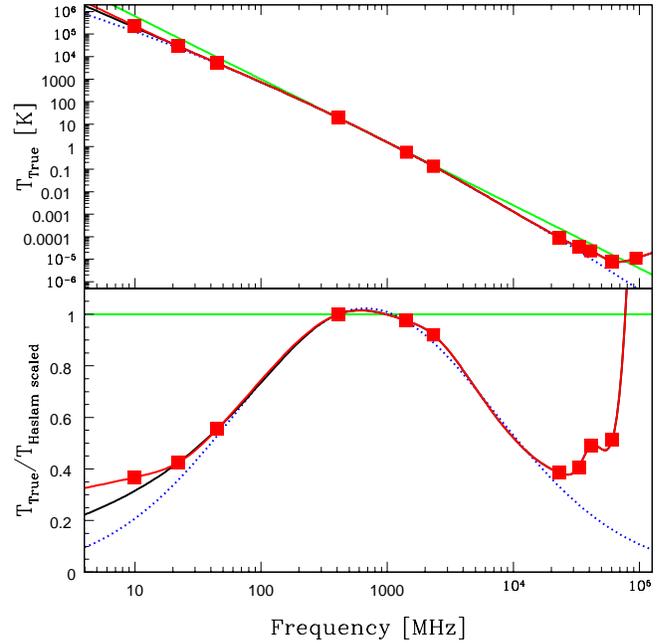}
\caption{
    Comparison between the different GSM methods presented in 
    \protect\Sec{secmethodcomparison}. Squares show the 11 
    measurements available at a pixel at $(l,b)$=(11.3$^{\circ}$,89.6$^{\circ}$). 
    Lines show fits based on 
    power-law-scaling of the Haslam map (straight solid/green line), 
    a quadratic polynomial in log-log (dotted/blue curve), 
    a cubic spline without the leftmost point (straight solid/black), 
    and a 3-component PCA fit (straight solid/red).
    In the lower panel, the curves have been divided by the 
    power-law-scaling of the Haslam map to make 
    discrepancies between the methods even more visible. 
    }
\label{MethodComparisonFig}
\end{figure}

\subsection{Method comparison}\label{secmethodcomparison}
\subsubsection{Single-component models}

The Galactic InterStellar Medium (ISM) is a highly complex medium 
with many different constituents interacting through a multitude of 
physical processes. Free electrons spiraling around the Galactic 
magnetic field lines emit synchrotron radiation \cite{rybicki}.
For the lower frequencies where synchrotron radiation is expected to 
dominate the Galactic emission, a common approach in the literature 
has been to simply scale the Haslam map \cite{haslam2} in 
frequency, usually with a power law 
\beq{HaslamPowerLawScalingEq}
   T(\rhat_i,\nu) = T(\rhat_i,\nu_*) \left(\frac{\nu}{\nu_*}\right)^{\beta}, 
\eeq
where $\rhat_i$ is the unit vector pointing toward the $i^{\rm th}$ pixel
in the map, $\nu$ is the frequency which this map is being scaled to, 
$\nu_*=408\>$MHz is the Haslam frequency, $\beta$ is the spectral 
index\footnote{The straight green line in the upper panel of 
     \protect\Fig{MethodComparisonFig} shows the case $\beta=-2.8$.}, 
and $T$ is the brightness temperature.   
However, the frequency dependence is known not to be a perfect power 
law: at higher frequencies, the slope of the synchrotron spectrum typically 
steepens\footnote{One expects a spectral
	steepening towards higher frequencies, corresponding to a 
	softer electron spectrum (\protect\cite{banday91}; Fig 5.3 in 
	\protect\cite{jonas99}). A recent analysis done at 22~MHz 
	\protect\cite{roger} shows that $\beta$ varies slightly over
	a large frequency range \cite{milo95,reich98a,reich98b,platania}.},
and other Galactic components such as free-free and dust emission begin to 
dominate. This suggests the use of a more general scaling of the type
\beq{HaslamScalingEq}
  T(\rhat_i,\nu) = T(\rhat_i,\nu_*)f(\nu_i),
\eeq	
where the spectrum $f(\nu_i)$ is optimized by fitting to maps at other 
frequencies. We will quantify the accuracy of this approach in \Sec{ResultsSec}. 
The main problem with it is that the foreground frequency dependence 
is known to vary across the sky. This occurs both because the synchrotron 
spectral index $\beta$ depends on the energy distribution of relativistic 
electrons \cite{rybicki}, which varies somewhat across the sky, and also 
because the ratio of synchrotron to dust and other emission components 
can vary from place to place. In contrast, \eq{HaslamScalingEq} assumes 
that a map of Galactic emission looks the same at all frequencies except 
for an overall change in amplitude.

\subsubsection{Polynomial and spline fitting}

Now that so much data is available, it is tempting to allow much more general fitting 
functions such as polynomials or cubic splines. We tested both of these approaches here 
and found that they gave their most accurate results when fitting in log-log (when fitting 
$\lg T$ as a function of $\lg\nu$ rather than using $T$ and/or $\nu$ directly), since 
the function to be fit is then rather smooth -- see \fig{MethodComparisonFig} (top panel). 
For instance, the quadratic polynomial fit
\beq{QuadraticPolynomialEq}
   \ln T(\rhat_i,\nu) = \alpha(\rhat_i) + \beta(\rhat) \ln\frac{\nu}{\nu_*} 
		     + \gamma(\rhat_i)\left(\ln\frac{\nu}{\nu_*}\right)^2 
\eeq	
generalizes \eq{HaslamPowerLawScalingEq} to a position-dependent ``running'' $\gamma$ 
of the spectral index $\beta$. For a given pixel $i$, let $m_i$ denote the number of 
surveys that have observed it ($6\le m_i\le 11$). Re-writing \eq{QuadraticPolynomialEq} 
in a matrix form, we obtain
\beq{pca1}
   \ybf = \Abf \xbf + \nbf,
\eeq	
where $\ybf$ is an $m_i$-dimensional vector that contains (the logarithm of) the 
temperatures at the $i^{th}$ pixel at the $m_i$ survey frequencies, $\Abf$ is an 
$m_i\times 3$ matrix that encodes the frequency dependence, and $\xbf$ is a 
3-dimensional vector that contains the $\alpha$, $\beta$ and $\gamma$ values at 
the $i^{th}$ pixel. The extra term $\nbf$ denotes noise in the broadest sense 
of the word, \ie, receiver noise, uncorrected offsets and calibration errors, and 
any other systematic effects or other non-sky signals present in the survey maps.
This is an overdetermined system of linear equations since we always have $m_i>3$ input 
maps available, and assuming that the noise has zero mean, \ie, $\expec{\n}=\zero$, 
the minimum-variance estimator for $\xbf$ is 
\beq{pca2}
   \xhat = \left[\A^t\N^{-1}\A\right]^{-1}\A^t\N^{-1}\y,
\eeq	
with covariance matrix
\beq{pca3}
   \NN = \left[ \Atbf \Nbf^{-1} \Abf \right]^{-1},
\eeq	
where $\Nbf$ is the noise covariance matrix $\expec{\nbf\nbf^t}$.
In \fig{MethodComparisonFig}, we have simply approximated $\Nbf$ 
by the diagonal matrix with $N_{ii}$ given by the $\rms$ of the $i^{th}$ map 
(we find the recovered maps to be rather insensitive to the choice of $\Nbf$). 
By performing this calculation for all the pixels in the sky, we obtain all-sky 
maps of the quantities $\alpha$, $\beta$ and $\gamma$. Finally, to obtain 
a sky map at any frequency $\nu$, we simply use these values of $\alpha$, 
$\beta$ and $\gamma$ in \eq{QuadraticPolynomialEq}.

We also tried the approach of fitting the (log-log) frequency dependence in 
each pixel to a separate cubic spline. This involves even more fitting parameters 
(between 6 and 11), as the resulting curve is forced to match the data exactly at 
all observed frequencies. Maps of $\alpha$, $\beta$ and $\gamma$ can then be 
produced by computing the first and second derivatives of the spline function.

\Fig{MethodComparisonFig} illustrates the pros and cons of the above-mentioned methods. 
Both the simple power law and the log-log quadratic polynomial are seen to provide 
poor fits simply because the physics is more complex than these functions can model. 
The figure shows that a power law is a poor approximation even in the synchrotron-dominated 
regime $\nu\simgt 1$~GHz, because of a distinct steepening of the spectrum towards higher
frequencies.
However, the figure shows that going to the opposite extreme and allowing too many 
fitting parameters causes problems as well, from overfitting the data. The spline blindly 
goes through all the data points without any regard for what constitutes physically 
reasonable behavior, and sure enough is seen to perform poorly when forced to 
extrapolate. The ability to extrapolate reliably is crucial to our sky model because 
many of our input maps have only partial sky coverage. A related drawback of the spline 
approach is that if one of the input maps has more noise or systematic errors than others,
this will fully propagate into the model rather than getting ``voted down'' by the other 
input maps.

A final problem, seem most clearly in the bottom panel, is caused by fitting the log of 
the temperature rather than the temperature itself: a relatively modest error in the 
predicted log-temperature translates into an exponentially amplified error in the 
temperature itself. The logarithmic fitting also complicates the modeling of measurement 
errors: if they are symmetrically distributed around zero and uncorrelated with the sky 
signal in the raw input maps, this is no longer true for the log-maps. In contrast, a 
linear combination of the linear input maps would preserve such desirable statistical 
properties of the noise.

\begin{figure} 
\includegraphics[width=9cm]{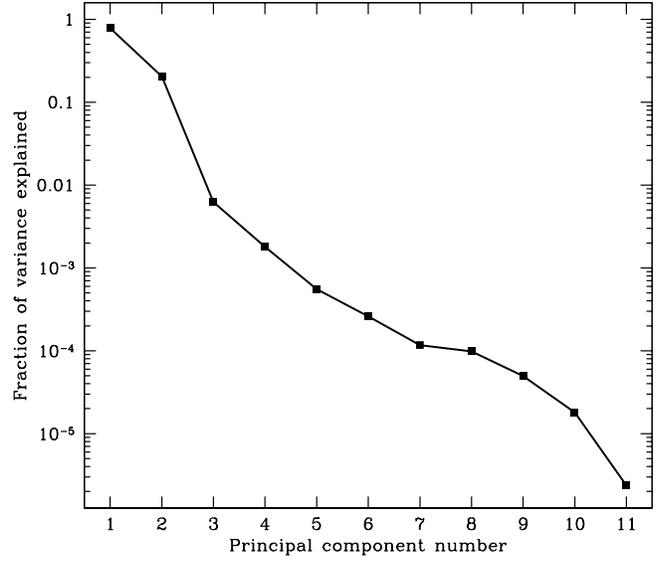}
\caption{The eigenvalues $\lambda_i/11$ for the 11 principal components, 
 	which can be interpreted as the fraction of the total variance 
	at the 11 frequencies that each component explains.
	}
\label{LambdaFig}
\end{figure}

\subsubsection{Principal Component Analysis} \label{PCAsec}

The above examples suggest that we should try a method that: 
 	{\it (1)} can fit the spectral behavior of the data with as few parameters as possible; and
 	{\it (2)} is linear (takes some linear combination of the raw input maps).
In other words, we want a linear fitting method where the data itself is allowed 
to select the optimal parametrized form for the frequency dependence.
Fortunately, the standard tool known as Principal Component Analysis (PCA) does 
exactly this \cite{NumericalRecipes}. Indeed, we find that PCA performs better 
than all the approaches tried above when we implement it as described below.

We begin by estimating the $11\times 11$ matrix of second moments
\beq{CdefEq}
	\C\equiv\frac{1}{\npix}\sum_{i=1}^{\npix}\y_i\y_i^t
\eeq
by averaging over all of the $\npix$ pixels $i$ that were observed in all 11 frequencies
(the sky region marked as ``123456'' in 
\fig{SurveysSkyCovarageFig})\footnote{If we had also 
	removed the mean of each map in this region 
	(an issue to which we return latter), $\C$ 
	would simply be the covariance matrix between 
	the 11 frequencies.}. 
Therefore, the quantities
\beq{RdefEq}
 	\sigma_j\equiv\C_{jj}^{1/2}
\eeq
simply give the rms fluctuations at each frequency, and the  correlation matrix 
\beq{RdefEq}
	\R_{jk}\equiv\frac{\C_{jk}}{\sigma_j\sigma_k}
\eeq
corresponds to the dimensionless correlation coefficients between all pairs of frequencies;
$-1\le \R_{jk}\le 1$ and $\R_{jj}=1$.
Defining the {\it normalized maps} $\z_i$
as the input maps rescaled to have rms fluctuations of unity at each frequency,
$\R$ is simply the matrix of second moments of these normalized maps.

\begin{figure} 
\includegraphics[width=9cm]{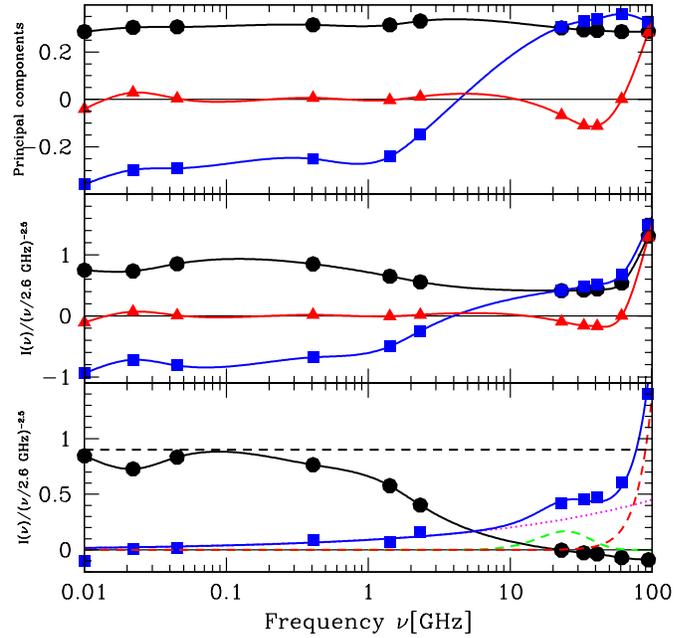}
\caption{The frequency dependence is plotted for the first three principal components, 
	labeled by black dots, blue squares and red triangles, respectively.
	The top panel is in units of the total rms fluctuations at each frequency, 
	whereas the middle panel shows the sky brightness temperature 
	divided by $(\nu/2.6\>$GHz$)^{-2.5}]$ to keep all frequencies on roughly 
	the same scale.
	The bottom panel shows
	typical spectra of various physical components \protect\cite{foregpars}:
	synchrotron $\propto\nu^{-2.5}$  (long-dashed black), 
	free-free $\propto\nu^{-2.15}$ (dotted magenta),
	spinning dust  (long-dashed green) and 
	thermal dust   (long-dashed red).
	It also shows half the sum (black dots) and difference (blue squares) of the first two components, 
	which are seen to behave roughly as 
	synchrotron (with a spectral index that steepens toward higher frequency) and a sum of free-free, spinning and thermal dust 
	(blue curve), respectively. 
	}
\label{ComponentFig}
\end{figure}

We then diagonalize the matrix $\R$, performing a standard eigenvalue decomposition 
\beq{EigenEq}
	\R = \P\LL\P^t,
\eeq
where $\P$ is an orthogonal matrix $(\P^t\P=\P\P^t=\I)$ whose columns are the 
eigenvectors (principal components) and $\LL_{jk}=\delta_{jk}\lambda_j$ is a diagonal 
matrix containing the corresponding eigenvalues, sorted in decreasing order.
The eigenvalues $\lambda_i$ are plotted in \fig{LambdaFig}, and the first three 
principal components are listed in \tab{ComponentTable} and shown in \fig{ComponentFig}. 
In this same table we also show the rms of the of each of the frequency maps calculated 
in the region 123456 (second column).

\begin{table}
\begin{center}
\caption{The three first principal components.}
\bigskip
{\footnotesize
\begin{tabular}{|l|r|c|r|r|}
\hline
 \multicolumn{1}{|c|}{$\nu$}	
& \multicolumn{1}{|c|}{rms}
&Comp1 &Comp2 &Comp3  \\ 
$[$GHz$]$&            &	       &	& 	\\
\hline
 0.010  & 262344~~~\,K    &0.286   & -0.358  &-0.121 \\
 0.022  &  33693~~~\,K    &0.304   & -0.297  & 0.086 \\
 0.045  &   6506~~~\,K    &0.306   & -0.291  & 0.010 \\
 0.408  &   25.4~~~\,K    &0.315   & -0.251  & 0.020 \\
 1.420  &  0.862~~~\,K    &0.314   & -0.242  &-0.008 \\
 2.326  &  0.204~~~\,K    &0.331   & -0.147  & 0.035 \\
23      &    0.541\,mK    &0.301   &  0.306  &-0.199 \\
33      &    0.230\,mK    &0.294   &  0.331  &-0.327 \\
41      &    0.140\,mK    &0.291   &  0.341  &-0.335 \\
61      &    0.065\,mK    &0.286   &  0.361  & 0.003 \\
94      &    0.053\,mK    &0.287   &  0.326  & 0.847 \\
\hline								        		 
\end{tabular}
}
\label{ComponentTable}
\end{center}
\end{table}

\begin{figure} 
\includegraphics[width=8.5cm]{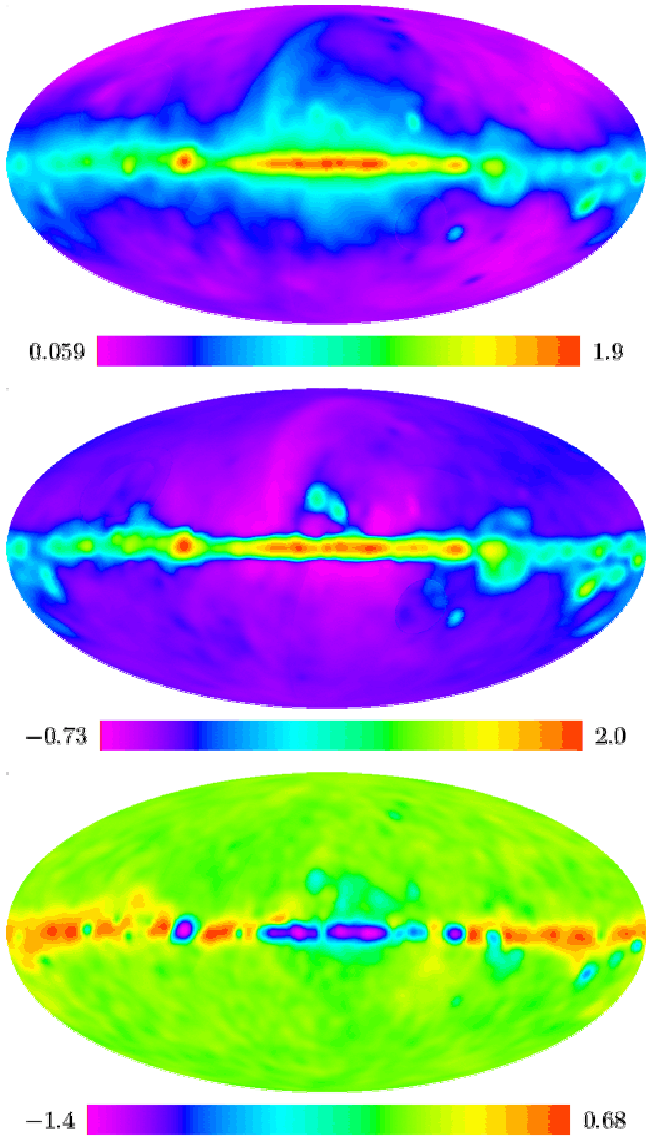}
\caption{The first three principal components, which can be crudely interpreted 
         as maps of total ``stuff" (top),
	 synchrotron fraction (middle)
	 and thermal dust fraction of non-synchrotron emission (bottom). 
	 The color scale corresponds to $\lg(T/1$K$)$ in the top panel, and 
	 $\sinh^{-1} (T/1$K$)/\ln 10 $ in the other panels to handle negative values 
	 (since $\sinh^{-1} x/\ln 10 \approx \lg |x|$ for $x\gg 1$ and
	 for large positive and negative values, while it is roughly linear near zero).
	 }
\label{ComponentMapFig}
\end{figure}

To help intuitively interpret this decomposition, \fig{ComponentMapFig} shows
maps of the first few principal components. Each principal component map $a_i$ is 
defined as the dot product of the corresponding eigenvector with the normalized 
multi-frequency vector $\z_i$ for each pixel. For each pixel $i$, we can therefore 
transform back and forth between the normalized multi-frequency vector $\z_i$ and 
the principal component vector $\a$ using the relations 
\beq{BackAndForthEq}
	\a_i=\P^t\z_i,\quad \z_i=\P\a.
\eeq
The principal component maps can be thought of as a division of the information 
in the 11 input maps into 11 mutually exclusive and collectively exhaustive chunks.
They are mutually exclusive in the sense that they are uncorrelated:
\beq{PCPeq}
	\expec{(\P^t\z)(\P^t\z)^t}=\P^t\expec{\z\z^t}\P=\P^t\C\P=\Lambda.
\eeq
They are collectively exhaustive in the sense that they together specify the 
multifrequency information completely through \eq{BackAndForthEq}. Moreover, 
\fig{LambdaFig} shows that almost all this information is contained in the 
first few principal components. Taking the trace of \eq{EigenEq} shows that
\beq{TraceEq}
	\sum_{j=1}^{11}\lambda_j = \tr\Lambda=\tr\Lambda\P^t\P=\tr\P\Lambda\P^t=\tr\R=11,
\eeq
since the diagonal elements of the correlation matrix are all unity.
In other words, the total variance to be explained in the normalized 
multifrequency data is 11, with a contribution of unity from each of 
the 11 normalized input maps, and \eq{PCPeq} shows that the $\jth$ 
principal component map explains a variance $\lambda_j$, \ie, a fraction 
$\lambda_j/11$ of the total.

\Fig{LambdaFig} shows that the first component (top panel of \fig{ComponentMapFig}) 
explains 80\% of this total variance, the second component explains another 19\%, 
the third explains another 0.6\%, and all the remaining eight components combined 
explain merely the last 0.3\%.
This is very convenient: we set out searching for a way to accurately parametrize 
the frequency dependence of the radio sky with as few parameters as possible, 
and have found that as few as two parameters capture more than 99\% of the information. 

Although principal component analysis is quite a standard data analysis technique
\cite{NumericalRecipes}, our analysis also includes some non-standard procedures, 
tailored for the particular challenges that our global sky modeling problem poses:
  \begin{itemize}
	\item We diagonalize $\R$ rather than $\C$.
	\item We perform no mean removal.
	\item We make up for missing data by fitting to only the best principal components.
	\item We perform frequency interpolation by splining $\lg\sigma_i$ and the 
	      component spectra.
  \end{itemize}
Let us now explain each of these procedures in more detail.

Diagonalizing $\R$ rather than $\C$ corresponds to using the normalized maps rather 
than the raw maps as input for the PCA. We made this choice because we are equally 
interested in providing a good fit (in terms of percent of rms explained) at all 11 
frequencies. If one took the raw maps as input, the PCA would instead focus almost 
entirely on optimizing the fit to the lowest frequency maps, since the increase of 
synchrotron temperature towards lower frequencies causes them to have by far the 
largest rms signal measured in Kelvin. This usage of the normalized maps also has 
the advantage that the spectra of the dominant physical components become rather 
gently varying functions of frequency, which makes them much easier to linearly 
fit (see \fig{ComponentFig}). This eliminates the need for logarithmic fits and their 
above-mentioned problems.

In a standard PCA, one diagonalizes the covariance matrix 
	$\expec{(\z-\expec{\z})(\z-\expec{\z})^t}$.
We instead diagonalize the matrix $\expec{\z\z^t}$, \ie, do not subtract off the mean 
from the normalized maps before computing their second moment matrix.
This is because, as quantified in \Sec{ResultsSec}, this procedure makes the method 
more accurate in regions with incomplete data: whereas the principal components from 
the region with 11 frequency data work well across the entire sky (basically, because 
they reflect underlying physical emission mechanisms which are the same everywhere), 
the 11 mean values from this region are not at all representative for other regions, 
as they depend strongly on Galactic latitude. By not removing the mean, we also exploit 
the physical property that none of the dominant foreground components can ever contribute 
a negative intensity\footnote{The only sky signal with a non-CMB spectrum that 
	can give a negative temperature contribution is the thermal SZ effect, 
	and it makes a rather negligible contribution compared to the synchrotron, 
	free-free and dust components.}.

Whereas a standard PCA can be performed in the region shown in \fig{SurveysSkyCovarageFig} 
where all 11 frequencies have been observed, we wish to build a global sky model covering 
the entire sky. Fortunately, we have $m_i\ge 6$ measured frequencies available everywhere, 
and have found that much fewer than 6 parameters are required for an excellent fit.
We therefore take the best $\mstar$ principal components determined in the region with 
complete data and fit them to the data available. In \Sec{ResultsSec} we will explore 
what is the best choice of $\mstar$ by quantifying the accuracy attained using $1\le \mstar\le 5$ 
components. We perform this fitting by modeling the observed data in a pixel with $m_i$ 
observed frequencies as
\beq{ModelingEq2}
	\z_i = \Pt_i\at_i +\n_i,
\eeq
where the tildes indicate that we are truncating to only $\mstar$ components:
$\Pt_i$ is the $m_i\times\mstar$-dimensional matrix containing the $\mstar$ first principal components as its columns,
$\at_i$ is the $\mstar$-dimensional vector corresponding to the first $\mstar$ principal component maps (see \fig{ComponentFig}), 
$\z_i$ contains the $m_i$ normalized input maps that have data for this pixel,
and the residual $\n_i$ models random noise from both measurement errors and additional 
principal components not included in the fit. We perform this fitting separately for each 
pixel $i$ by minimizing
\beq{chi2Eq}
	\chi^2\equiv (\z_i-\Pt_i\at_i)^t\N_i^{-1}(\z_i-\Pt_i\at_i),
\eeq
which gives the solution
\beq{BestFitEq2}
	\at_i = \left[\Pt_i^t\N_i^{-1}\Pt_i\right]^{-1}\Pt_i^t\N_i^{-1}\z_i.
\eeq	
We describe our choice for the ``noise'' covariance matrix $\N_i\equiv\expec{\n\n^t}$ 
in \Sec{ResultsSec}.

\begin{table*}
\begin{center}
\caption{Relative {\it rms} error in the sky region 123456.}
{\footnotesize
\begin{tabular}{|l|c|c|c|c|c|c|c|}
\hline
\hline
 \multicolumn{1}{|c|}{$\nu$}	
&Optimal &  \multicolumn{5}{|c|}{Principal components used}	&Unexplained\\
$[$GHz$]$ & 	 &1 	  &2 	   &3	    &4	     &5  	&fraction\\
\hline  
 0.010  &0.062  &0.543   &0.078   &0.072   &0.065   &0.066   &0.00387  \\
 0.022  &0.036	&0.450   &0.064   &0.060   &0.039   &0.038   &0.00126  \\
 0.045  &0.035	&0.438   &0.046   &0.046   &0.038   &0.038   &0.00121  \\
 0.408  &0.034	&0.379   &0.044   &0.044   &0.039   &0.039   &0.00115  \\
 1.420  &0.111	&0.386   &0.135   &0.135   &0.150   &0.196   &0.01235  \\
 2.326  &0.075	&0.235   &0.084   &0.083   &0.081   &0.137   &0.00562  \\
23      &0.015  &0.463   &0.058   &0.026   &0.026   &0.026   &0.00024  \\
33      &0.006  &0.504   &0.086   &0.009   &0.008   &0.008   &0.00004  \\
41      &0.009	&0.519   &0.089   &0.017   &0.015   &0.015   &0.00008  \\
61      &0.018	&0.542   &0.023   &0.023   &0.023   &0.023   &0.00033  \\
94      &0.057	&0.538   &0.225   &0.059   &0.059   &0.059   &0.00328  \\
\hline			     	 							    
\hline
\end{tabular}
}
\label{AccuracyTable1}
\end{center}
\end{table*}

Let us summarize the above steps: we first find the principal components using the sky 
region with data at all 11 frequencies, then use the frequency dependence of these best 
$\mstar$ components to fit for maps of their amplitudes across the entire sky.
This leaves us with an all-sky model predicting the emission at the 11 frequencies.
However, we also wish to predict the emission at {\it any} frequency between $10\>$MHz$\le\nu\le 100\>$GHz. 
We do this by fitting the frequency dependence of   both $\lg\sigma_j$ and each of
the $\mstar$ principal components with a cubic spline as a function of $\lg\nu$.
This works well only because, as seen in \fig{ComponentFig}, these are smooth, slowly 
varying functions. In contrast, it is notoriously difficult to perform useful interpolation 
of  matrices, \eg, $\R$, without wreaking havoc with their eigenvalues and physical
behavior.


\section{Results}\label{ResultsSec}

Above we have described how we construct our GSM.
However, how accurate is it, and what does it teach us?

\subsection{Accuracy of our GSM}
\subsubsection{Accuracy in the fully mapped region}

\begin{figure*} 
\includegraphics[width=18.3cm]{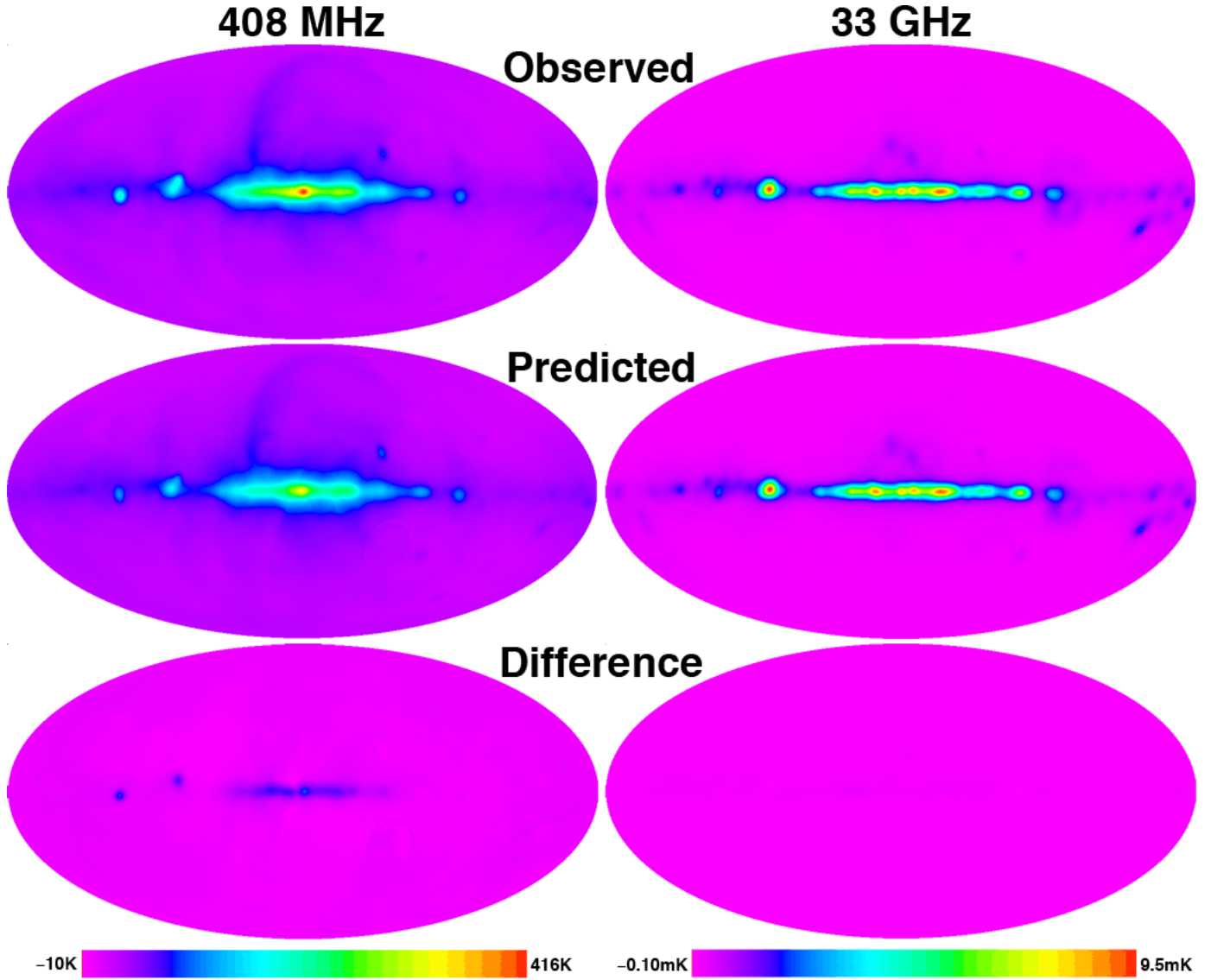}
\caption{Accuracy of our model at 408 MHz (left column) and 33 GHz (right column).
	 The top row shows the observed data (the Haslam and WMAP Ka-band maps),
	 the middle row shows the maps predicted by our 3-component GSM without 
	 using the observed map above it, and the bottom row shows the observation 
	 minus the prediction (which is visually indistinguishable from zero for 
	 the 33 GHz case, because the residuals are less than $1\%$ in and around the Galactic plane).
         }	
\label{PredictionFig}
\end{figure*}

\Tab{AccuracyTable1} shows the accuracy of our GSM in the sky region where 
we have data at all 11 frequencies. As described in \Sec{AccuracyCriterionSec},
we measure how accurately each map can be predicted by the other maps.
Specifically, for each of the 11 frequencies, we compute the difference 
between this map and the map predicted by using only information from the 
other 10 maps, then tabulate the rms of this difference map divided by the 
rms of the observed map. \Fig{PredictionFig} illustrates this procedure 
for the more challenging all-sky case to which we return below: for these two example, 
the relative rms error is the rms of the bottom panel divided by the rms 
of the corresponding top panel.

Not surprisingly, \Tab{AccuracyTable1} shows that using two components 
is much more accurate than using only one, reducing errors by almost an order
of magnitude at some frequencies. Adding a third component is seen to further
improve the accuracy, although not by as much, and mainly in the 20-40 GHz range. 
Adding a fourth component produces only minor gains, and reduces the 94 GHz 
accuracy ever so slightly, and adding a fifth component makes the accuracy 
noticeably worse at three frequencies, suggesting that we are beginning to 
overfit the data just as with the above-mentioned cubic spline approach.

It is interesting to compare these accuracy numbers with what information
theory tells us is the best possible case. The most accurate prediction 
for a given map using a linear combination of the others is easily computed 
using a standard linear regression analysis \cite{NumericalRecipes}, and 
these optimal errors are listed in the second column of \Tab{AccuracyTable1}.
A popular statistical measure of how useful something is for predicting 
something else is the fraction of the variance that it explains. 
Under the heading of ``unexplained fraction'', we therefore also tabulate
the fraction $f_j$ of the map variance that is not explained by the other maps; 
this is simply the square of the rms residual, since the maps are normalized 
to have total variance of unity. Note that there is no need to actually 
perform a linear regression to compute these optimal numbers, as a simple 
derivation shows that they can be computed directly from the correlation matrix: 
\beq{UnexplainedEq}
	f_j = \frac{1}{(\R^{-1})_{jj}\R_{jj}}.
\eeq

The results of this analysis are very encouraging. \Tab{AccuracyTable1} shows 
that the residuals achieved by our GSM with three components are very close 
to these smallest possible ones, which means that we need not worry about 
having overlooked some alternative modeling method that does much better.
The results also raise an important question:
	if linear regression is so good, why do we not use it instead of our GSM?
The answer is that we cannot: regression only works when the matrix $\R$ is known, 
and we can only compute $\R$ when we already have data at the frequency that 
we are trying to model. In other words, whereas we can use regression for accuracy 
testing, where we already know the answer, it does not help us with modeling all 
the unobserved frequencies between 10 MHz to 100 GHz. We did explore the idea of 
predicting the $\R$-matrix entries corresponding to new frequencies using 
interpolation, but were unable to obtain useful results. In contrast, our GSM is 
straightforward to interpolate to other frequencies, because we simply need to 
interpolate the spectra plotted in \fig{ComponentFig}. 

When we presented our GSM method described in \Sec{MethodSec}, there were two 
details that we never specified: the choice of noise covariance matrix $\N$ 
in \eq{BestFitEq2} and the choice of $\mstar$, the number of principal components 
to use. Let us now discuss these two choices in turn.

\subsubsection{The noise covariance matrix}

The ``noise'' is simply the residual signal in a map that we are unable to predict using the 
other maps, so it will contain contributions from both measurement errors in the input maps and 
   sky emission mechanisms modeled with inadequate precision. 
Both of these contributions are captured by the remaining 
principal components not included in the fit, which according to \eq{EigenEq} 
make a contribution to $\R$ that is $\P\LL\P^t$ except with all eigenvalues from 
the included components set to zero.
However, it is easy to show that adding noise for the included components has no 
effect on the solution of \eq{BestFitEq2}, so we get exactly the same result if 
we simply set $\N=\P\LL\P^t=\R$. As a reality check, we also tried the alternative 
approach of setting $\N$ equal to a diagonal matrix with the optimal variance values 
from \tab{AccuracyTable1} on the diagonal, and obtained very similar results.

\subsubsection{Accuracy across the entire sky}

\begin{table}
\begin{center}
\caption{Relative error averaged over the entire sky}
{\footnotesize
\begin{tabular}{|l|c|c|c|c|c|}
\hline
\hline
 \multicolumn{1}{|c|}{$\nu$}	
&\multicolumn{5}{|c|}{Principal components used}\\
$[$GHz$]$ &1 	 &2 	  &3	   &4	   &5  \\
\hline 
 0.010  &0.438   &0.091   &0.098   &0.088   &0.168 \\
 0.022  &0.690   &0.164   &0.144   &0.142   &0.138 \\
 0.045  &0.712   &0.110   &0.109   &0.111   &0.100 \\
 0.408  &0.436   &0.112   &0.115   &0.127   &0.134 \\
 1.420  &0.546   &0.143   &0.144   &0.698   &0.875 \\
 2.326  &0.216   &0.148   &0.155   &0.158   &0.503 \\
23      &0.423   &0.082   &0.062   &0.062   &0.064 \\
33      &0.453   &0.077   &0.013   &0.013   &0.014 \\
41      &0.458   &0.071   &0.032   &0.032   &0.031 \\
61      &0.444   &0.069   &0.068   &0.068   &0.070 \\
94      &0.385   &0.223   &0.121   &0.121   &0.160 \\
\hline								        		 
\hline
\end{tabular}
}
\label{AccuracyTable2}
\end{center}
\end{table}

\begin{table*}
\begin{center}
\caption{rms sky signal in K for regions of different cleanliness}
{\footnotesize
\begin{tabular}{|l|c|c|c|c|c|c|}
\hline
\hline
 \multicolumn{1}{|c|}{$\nu$}	
&\multicolumn{6}{|c|}{($\leftarrow$cleaner)\quad region\quad (dirtier$\rightarrow$)}\\
$[$GHz$]$ &1 	&2     &3      &4     &5      &6     \\
\hline  
 0.010  &203740          &272337          &304115          &328310          &281838          &  	      \\ 
 0.022  & 27336          & 41972          & 63535          & 98153          &118713          &130600	      \\ 
 0.045  &  5486          &  8347          & 13019          & 21285          & 31287          & 35926 	      \\ 
 0.408  &    20.0        &    30.0        &    52.0        &   103.3        &	182.9	     &   230.4	      \\ 
 1.420  &     0.744      &     1.021      &	1.614	   &	 3.016	    &	  5.356	     &     6.839      \\ 
 2.326  &     0.150      &     0.238      &	0.487	   &	 1.184	    &	  2.196	     &     2.760      \\ 
23      &     0.000098   &     0.000260   &	0.001106   &	 0.004140   &	  0.010357   &     0.015078   \\
33      &     0.000036   &     0.000097   &	0.000435   &	 0.001693   &	  0.004343   &     0.006444   \\
41      &     0.000021   &     0.000056   &	0.000255   &	 0.000996   &	  0.002569   &     0.003851   \\
61      &     0.000007   &     0.000024   &	0.000117   &	 0.000433   &	  0.001091   &     0.001639   \\
94      &     0.000006   &     0.000022   &	0.000106   &	 0.000332   &	  0.000758   &     0.001078   \\
\hline								        		 
\hline
\end{tabular}
}
\label{rmsTable}
\end{center}
\end{table*}

 How many principal components should we include to maximize the GSM accuracy?
To determine this, we must quantify the accuracy not only in the best-case sky 
region where we have complete data, but also over the rest of the sky as well, since 
we ultimately care about the whole sky. \Tab{AccuracyTable2} shows how this 
sky-averaged accuracy depends on the number of components used. 
Specifically, we have computed the relative rms error just as in \tab{AccuracyTable1}, 
but separately for each of the 10 sky regions show in \fig{SurveysSkyCovarageFig}, 
then computed their average weighting by sky area. The numbers show that  
the two best choices are 3 and 4 components.
However, whereas these two choices were essentially tied in the fully observed region, 
$\mstar=3$ comes out slightly ahead in the all-sky average because it is twice as
accurate at 1.42 GHz.
This means that, although the 3-component model is slightly less sophisticated and 
therefore typically slightly less accurate, it is also more robust and less likely 
to go badly wrong in unusual parts of the sky. For this reason, we focus on the 
3-component model in the rest of this paper. 

\Tab{AccuracyTable2} shows that the typical accuracy is around 10\% for the frequencies 
below WMAP, and noticeably better for the four lowest WMAP-frequencies.
It is striking that the 33 GHz accuracy is as good as 1.3\%, which means that if WMAP 
had not made this particular map, as much as 99.97\% of the sky variance at this frequency 
could have been predicted by the other maps (and this is not even counting the CMB
signal, which we subtracted off at the outset). Also, as shown in the results for the 
WMAP 94 GHz map, the amount of noise in a map clearly worsens the accuracy to which 
we can reconstruct it.

\begin{figure} 
\includegraphics[width=9cm]{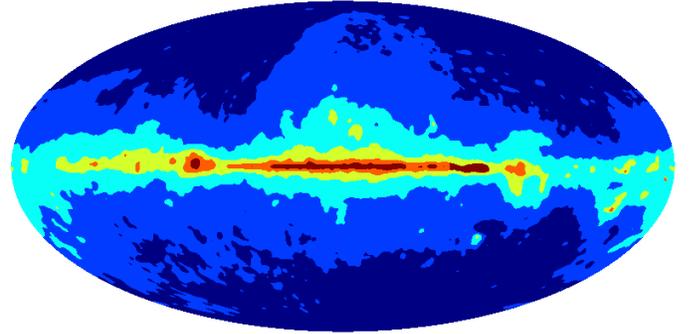}
\caption[1]{\label{CleanlinessFig}
	Our subdivision of the sky into six regions 
	of decreasing cleanliness. From outside in, they correspond to WMAP-based junk map 
	temperatures $T<100\uK$, $100\uK-300\uK$, $300\uK-1\mK$, $1\mK-3\mK$, 
	$3\mK-10\mK$, and $T>10\mK$, respectively.
	}
\end{figure}

\subsubsection{Accuracy at different Galactic latitudes}

It is difficult to quantify the accuracy of our GSM in a meaningful way with a single number, 
since the sky signal varies so dramatically with Galactic latitude: any measure of {\it absolute} 
error (in Kelvin) will therefore be dominated by the inner Galactic plane, while any measure of 
{\it relative} error will tend to be dominated by the cleanest regions where the signal-to-noise 
ratio is the poorest.
To get a more nuanced picture of how accurate our GSM is, let us therefore quantify the relative 
errors separately for the
regions shown in \fig{CleanlinessFig}, which subdivide the sky into six parts of increasing Galactic 
emission. They were defined in \cite{mapforegs} by computing the four differences of WMAP maps at 
neighboring frequencies (to subtract out the CMB), computing the largest absolute value at each pixel, 
and making a contour plot of the resulting ``junk map''.
From outside in, the regions correspond to junk map temperatures $T<100\uK$, $100\uK-300\uK$, 
$300\uK-1\mK$, $1\mK-3\mK$, $3\mK-10\mK$, and $T>10\mK$, respectively, so the typical sky signal 
differs by about half an order of magnitude between neighboring regions.
\Tab{rmsTable} shows that this scaling is roughly valid at all the WMAP frequencies, and that the 
differences between dirty and clean regions become less extreme towards lower frequencies.
Although we of course do not have complete frequency coverage across the entire sky, comparing 
\fig{SurveysSkyCovarageFig} with \fig{CleanlinessFig} shows that we are lucky to have 
coverage at all frequencies somewhere within each of the six sky regions of \fig{CleanlinessFig},
with the only exception that the very dirtiest region is not observed at 10 MHz.

\begin{table}
\begin{center}
\caption{Relative error as a function of sky cleanliness}
{\footnotesize
\begin{tabular}{|l|c|c|c|c|c|c|}
\hline
\hline
 \multicolumn{1}{|c|}{$\nu$}
&\multicolumn{6}{|c|}{($\leftarrow$cleaner)\quad region\quad (dirtier$\rightarrow$)}\\
$[$GHz$]$&1 	&2     &3      &4     &5      &6    \\
\hline  
 0.010  &0.091  &0.099  &0.094  &0.148  &0.111  &      \\
 0.022  &0.080  &0.082  &0.136  &0.210  &0.306  &0.451 \\
 0.045  &0.094  &0.102  &0.095  &0.108  &0.158  &0.202 \\
 0.408  &0.084  &0.088  &0.072  &0.125  &0.170  &0.190 \\
 1.420  &0.187  &0.133  &0.144  &0.180  &0.157  &0.129 \\
 2.326  &0.147  &0.160  &0.158  &0.165  &0.170  &0.167 \\
23      &0.111  &0.070  &0.083  &0.073  &0.062  &0.050 \\
33      &0.062  &0.022  &0.013  &0.012  &0.011  &0.011 \\
41      &0.091  &0.052  &0.042  &0.039  &0.034  &0.028 \\
61      &0.270  &0.067  &0.071  &0.082  &0.079  &0.067 \\
94      &0.766  &0.164  &0.095  &0.124  &0.136  &0.135 \\
\hline								        		 
\hline
\end{tabular}
}
\label{AccuracyTable3}
\end{center}
\end{table}

\Tab{AccuracyTable3} summarizes how the accuracy of our 3-component GSM depends on both 
frequency and Galactic signal level. At the sub-GHz frequencies relevant to 21cm tomography, 
we see that the accuracy is typically $\sim 10\%$ or better in the cleanest parts of the sky, 
and degrades in the inner Galactic plane.
For the higher frequencies relevant to CMB research, the situation is the opposite: 
the accuracy is best in the dirtiest parts of the sky (as good as $1\%$ at 33 and 41 GHz), 
but degrades in the cleanest regions. This is clearly due to the fact that detector noise 
is non-negligible at the higher WMAP frequencies, so that the lower the signal is, the 
lower the signal-to-noise level and the accuracy. Future WMAP data releases are therefore 
likely to further improve the accuracy of our GSM at CMB frequencies.

Finally, it is important to remember that the errors in our downloadable GSM are likely to be 
even smaller than the tables above suggest, because a map used as ``truth'' in a test may 
itself have noise and systematic errors, and also because it uses all 11 input maps jointly, 
not merely 10 at a time. For example, one can clearly make vastly better predictions 
near 408 MHz than \tab{AccuracyTable3} suggests if the Haslam map is included in the modeling.

\subsection{Implications for our input maps}

An interesting byproduct of our modeling effort is an independent quality assessment of 
the 11 input maps. If two maps are highly correlated, this implies that none of them can be 
afflicted by large noise or systematic errors, which would have spoiled the correlation.
More quantitatively, the unexplained variance fraction listed in \tab{AccuracyTable1} 
places an upper bound on the total contribution from detector noise and systematic 
errors in a map. If we focus on its square root, the optimal rms column in the same table, 
we see that the lowest frequency WMAP maps give the lowest residuals. This is not surprising,
considering that in order to meet its CMB science goals, WMAP had to be designed with 
significantly stricter systematic error control than typical in radio astronomy. As mentioned 
above, the WMAP increase in residuals with frequency reflects the drop in foreground signal while 
detector noise remains important and roughly constant.

At the lower frequencies relevant to 21 cm tomography, we see that the 10-408 MHz map errors 
can be at most at the 10\% level in the cleaner parts of the sky (see \tab{AccuracyTable3}), 
and no more than $6\%$ in the region where we have full frequency 
coverage (see \tab{AccuracyTable1}, column 2).
The remaining radio maps (at 1.42 GHz and 2.326 GHz)  
have error bounds about a factor two higher.
%

Finally, there is one kind of systematic error that our modeling cannot detect: an overall 
position-independent calibration error in a map. Because this would not affect the dimensionless 
correlation coefficients with other maps, it would not affect our goodness-of-fit either, merely 
cause corresponding calibration errors in the predictions.


\subsection{Physical interpretation of our GSM}
           
The goal of this paper is simply to model the Galactic emission, not to understand it physically. 
However, since our statistical results automatically encode interesting physical information,
let us briefly comment on possible interpretations.   
	  
\subsubsection{Component interpretation}

A number of physical components of Galactic emission in our frequency range have been 
thoroughly discussed in the literature, notably synchrotron radiation, free-free emission, 
spinning dust and thermal dust. However, we should not expect these physical components, 
which tend to be highly correlated, to match our principal components, which are by 
definition uncorrelated. 
We should instead expect our first principal component (top panel in \fig{ComponentMapFig}) 
to trace the total amount of ``stuff'', and the remaining principal components to describe 
how the ratios of different physical components vary across the sky.
The frequency dependence seen in \fig{ComponentFig} confirms this. The first component is shown 
to contribute an essentially constant fraction of the rms at all frequencies, corresponding 
to $\lambda_1/11\approx 80\%$ of the total variance.

The second component, which explains another $\lambda_2/11\approx 19\%$ of the total variance, 
is seen to have the negative of a synchrotron-like 
spectrum below a few GHz, and a spectrum at higher frequencies that is suggestive of a sum 
of free-free emission, spinning dust and thermal dust. This suggests that this component encodes what fraction of the total
emission is due to synchrotron radiation.
Sure enough, the second panel in \fig{ComponentMapFig} is seen to be negative in the north
polar spur region which is known to be dominated by synchrotron emission, and positive in 
regions like the inner Galactic plane and the Large Magellanic Cloud where 
one expects higher fractions of non-synchrotron emission.

The third component, which explains two thirds of the remaining variance (and $\lambda_3/11\approx 0.6\%$ 
of the total variance), is seen in \fig{ComponentFig} to have a spectrum that looks like thermal 
dust at the high end, goes negative below that, and essentially vanishes
below a few GHz where synchrotron radiation becomes dominant. This suggests crudely interpreting it as encoding what fraction of the
non-synchrotron signal is due to to thermal (vibrational) dust emission. 
It is unclear whether the 10 MHz blip in its spectrum is a 
fluke or reflects a physical correlation between dust properties and low-frequency synchrotron
properties like self-absorption \cite{petersonwebber2002}.

\begin{figure*} 
\includegraphics[width=17.5cm]{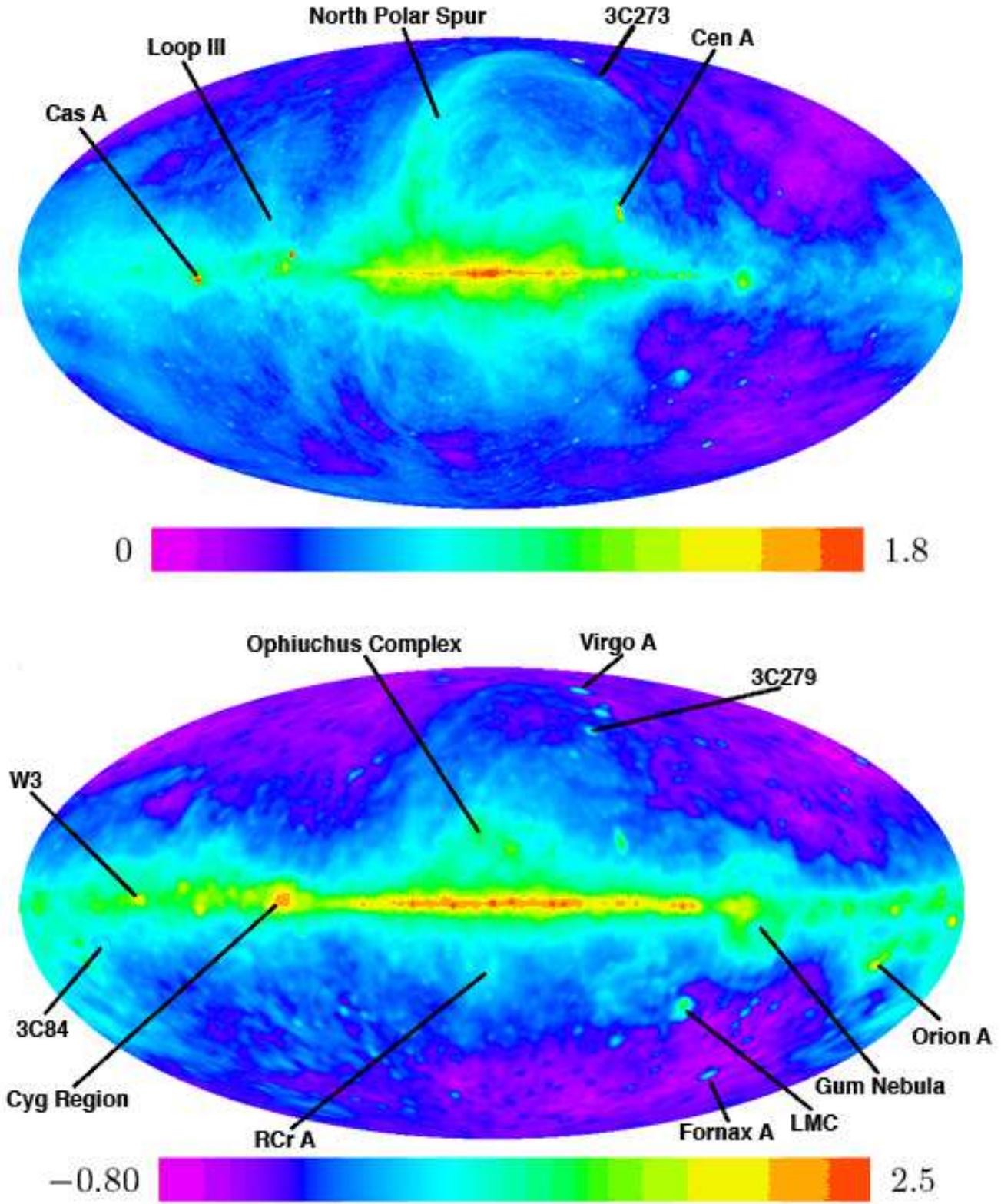}
\caption{
         Our synchrotron (top) and non-synchrotron (bottom) templates are the sum and difference of our first
	 two principal components, where the color scales corresponds to $\lg(T/1$K$)$.
	 Labels indicate bright objects in our Galaxy such as 
	 supernova remnants (Cas A, North Polar Spur, Loop III), 
	 an emission nebula (Gum Nebula), 
	 giant molecular clouds (Orion A, R Corona Australis, the Ophiucus Complex, W3) and an
	 active star-forming region (Cygnus Region)
	 as well as bright extragalactic sources like
	 giant elliptical galaxies (Virgo A, Fornax A),
	 radio galaxies (Centaurus A, 3C84) and
	 quasars (3C273, 3C279).
	 }
\label{blobsFig}
\end{figure*}

\subsubsection{Synchrotron and non-synchrotron templates}

As we discussed before, 
we do not expect our principal components to correspond directly to physical components, because the former are by definition
uncorrelated while the latter are not (``stuff traces stuff'', and there tends to be more of everything at low Galactic latitudes).
However, it is interesting to ask whether we can form linear combinations of our principal components that have a simple 
physical interpretation.

Interestingly, we can. 
In \fig{ComponentFig}, we see that taking the sum and difference of the first two principal components (from the second panel)
gives components whose spectra look distinctly like what is theoretically expected for synchrotron and a combination 
of the other emission components, respectively (as seen in the bottom panel).
First of all, \fig{ComponentFig} (bottom) shows that the two new template spectra are approximately non-negative at all 11 frequencies. 
This is a non-trivial result, since generic 11-dimensional 
eigenvectors or combinations of them will have both significantly negative and significantly positive components --- in contrast, we know that
neither synchrotron, free-free nor dust emission can be negative.
Second, the same figure shows that first template, which we will hereafter refer to as our synchrotron template, 
has a spectral index 
$\beta\approx -2.5$ at low frequencies, gradually steepening towards higher frequencies just as expected 
for synchrotron radiation \cite{banday91,jonas99}.
In contrast, the second template, which we will refer to simply as our non-synchrotron template, 
is seen to have a spectrum such that $\nu^{2.5}I(\nu)$ rises toward higher frequencies, and can be fit by a sum of 
free-free, spinning dust \cite{DL98} and thermal dust emission.
The corresponding sky maps (the sums and differences of the first two principal components) are shown in \fig{blobsFig}.

\begin{figure*} 
\includegraphics[width=18.3cm]{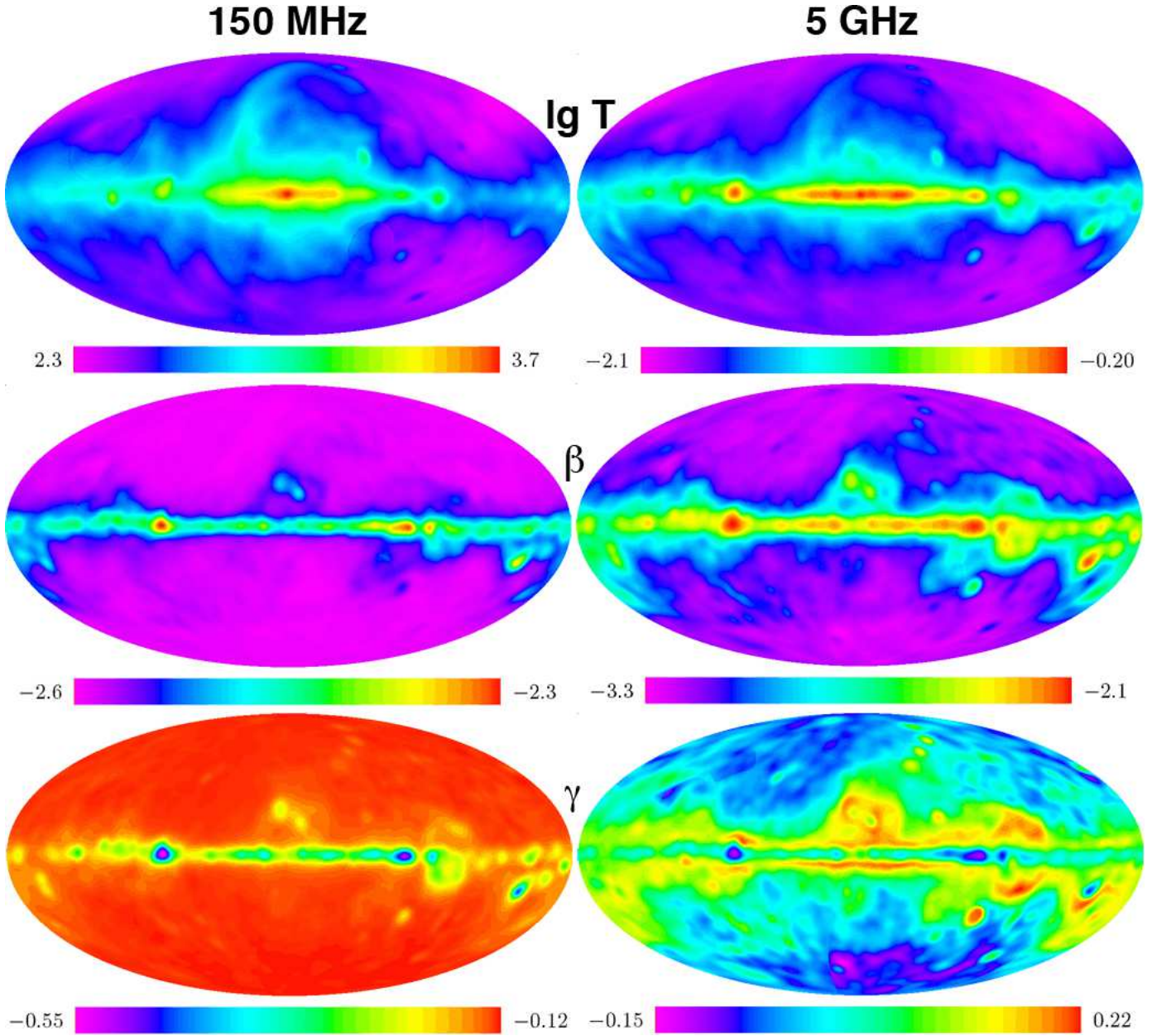}
\caption{
         Sky maps (from top to bottom) of the temperature, spectral index $\beta$, 
	 and "the running" (or variation) of spectral index $\gamma$ at 150 MHz (shown on the left) and
	 5 GHz (shown on the right).
 	 Whereas the 150 MHz emission is dominated by synchrotron radiation with 
 	 a spectrum that is both falling ($\beta\sim -2.5$) and steepening ($\gamma<0$),
 	 the 5 GHz emission has a much broader range of spectral indices 
	 that are mostly getting less negative towards higher freqency ($\gamma>0$). 
	 }
\label{AlphaBetaGammaFig}
\end{figure*}

In other words, our spectral results are consistent with the interpretation that the top panel of \fig{blobsFig}
is a diffuse synchrotron template while the bottom one is a template of diffuse non-synchrotron emission.
As expected, known supernova remnants subtending large angles (Cas A, North Polar Spur and Loop III)
are prominent in the synchrotron template, while diffuse dusty sources like the Cygnus Region stand out in the other template.
However, we cannot make any such interpretations for the point sources that appear in the paper.
This is because some point sources were removed from some of the low-frequency radio maps we used,
which can fool our analysis into removing them from the synchrotron template.
For example, in the 22 MHz map that we used, areas around the strong point sources Tau A, Cas A, Cyg A and Vir A.
have been blanked. At 1420 MHz, only Cas A was blanked.
Although it would be useful to repeat our analysis with new versions of the input maps where point sources have not been removed
(or where they have been reinserted using measured fluxes), the present paper of course has the opposite focus:
our key purpose is to model the diffuse emission for use in the {\it cleanest} parts of the sky, 
which are the ones most relevant to 21 cm tomography and CMB observations.

It is worth emphasizing the blind nature of our analysis: by simply forming those two unique linear combinations of our two dominant 
principal components for which the spectra were non-negative, our approach discovered the synchrotron and non-synchrotron spectra 
in the data using no physics input whatsoever.

\subsection{Angular resolution options}

To be able to use all 11 of our input maps, our spectral modeling has been performed at 
$5.1^\circ$, the lowest common denominator.
If we make the approximation that the spectral shape, but not its amplitude, varies only slowly 
across the sky, then we can create a higher resolution global sky model by locking the 
amplitude to a higher resolution input map. 
For example, for each pixel, we can rescale all three principal components used by the same constant, chosen
such that the prediction at 408 MHz matches the full resolution Haslam map. 
This procedure is illustrated in \fig{blobsFig}: the top panel locks to the $1^\circ$ Haslam map (recommented for applications
below 1 GHz where synchrotron dominates) while the bottom panel locks to the WMAP 23 GHz map smoothed to $2^\circ$ 
to suppress detector noise (recommended for applications at CMB frequencies).
These $1^\circ$, $2^\circ$ and $5.1^\circ$ versions of our GSM are all available on the above-mentioned website.
\Fig{AlphaBetaGammaFig} shows examples of our output maps at $5.1^\circ$ resolution.

 
\section{Conclusions}

We have presented a global sky model for 10 MHz to 100 GHz Galactic emission derived from 
all publicly available total power large-area radio surveys, digitized with optical character 
recognition when necessary and compiled into a uniform format. Both our data compilation 
and software for returning a predicted all-sky map at any frequency from 10 MHz to 100 GHz 
are available at {\bf http://space.mit.edu/home/angelica/gsm}.

We found that a PCA-based model with only three components can fit the 11 most accurate data sets 
(at 10, 22, 45 \& 408 MHz and 1.42, 2.326, 23, 33, 41, 61, 94 GHz) to an accuracy around 1\%-10\% 
depending on frequency and sky region. We found that using these three principal components 
comes very close to the maximal accuracy allowed by information theory, with the added advantage 
of allowing robust frequency interpolation and some physical interpretation. The fact that our 
model has so few fitting parameters in a given spatial direction  also makes it rather robust 
to the input data: a map with lots of noise or systematic errors will have smaller correlations 
with other maps, and therefore and get ``voted down'' by the other maps and given less weight. 

Strong correlations between different physical emission mechanisms 
would explain why such accurate fits are possible with fewer principal components than known 
physical components: one rapidly counts beyond three when including free-free emission, spatial 
variations of the synchrotron and dust spectra, etc.


We have focused entirely on unpolarized Galactic emission. To help maximize the future scientific 
impact of 21 cm tomography experiments, it will be important to extend this work to both
extragalactic point sources and polarized emission. Since these experiments will provide a gold 
mine of cosmological information buried by under $\sim 10^4$ times larger foreground signals,
this should be well worth the effort!

\section*{Acknowledgements}

We would like to thank Ben Gold, Jacqueline Hewitt, Miguel Morales, 
Wolfgang Reich and Matias Zaldarriaga for helpful comments, and 
Hector Alvarez and Koitiro Maeda for generously allowing us to use 
the 45 MHz radio survey.
We thank the the WMAP team for making their data public via
the Legacy Archive for Microwave Background Data 
Analysis (LAMBDA) at {\bf http://lambda.gsfc.nasa.gov}.
Support for LAMBDA is provided by the NASA Office of Space Science.
We thank Krzysztof G\'orski and collaborators for creating the 
HEALPix package \cite{healpix}.
This work was supported by NSF grants AST-0607597, 0134999 and 6915954, 
by NASA grant NAG5-11099, 
by the Kavli Foundation, 
by the fellowships from the David and Lucile Packard Foundation and the Research Corporation, and
support from the Australian Research Council through Federation Fellowship FF0561298.


\label{lastpage}
\end{document}